\documentclass[%
reprint,
superscriptaddress,
showpacs,
preprintnumbers,
 bibnotes,
 amsmath,amssymb,
 aps,
prl
]{revtex4-1}

\usepackage{graphicx}
\usepackage{dcolumn}
\usepackage{bm}
\usepackage{color}

\begin{document}


\title{Doping-dependent phonon anomaly and charge-order phenomena in the HgBa$_{2}$CuO$_{4+\delta}$ and HgBa$_{2}$CaCu$_{2}$O$_{6+\delta}$  superconductors}

\author{Lichen~Wang}
\altaffiliation[Present address: ]{Max Planck Institute for Solid State Research, D-70569 Stuttgart, Germany}
\affiliation{International Center for Quantum Materials, School of Physics, Peking University, Beijing 100871, China}
\author{Biqiong~Yu}
\affiliation{School of Physics and Astronomy, University of Minnesota, Minneapolis, Minnesota 55455, USA}
\author{Ran~Jing}
\altaffiliation[Present address: ]{Department of Physics, Columbia University, New York 10027, USA}
\affiliation{International Center for Quantum Materials, School of Physics, Peking University, Beijing 100871, China}
\author{Xiangpeng~Luo}
\altaffiliation[Present address: ]{Department of Physics, University of Michigan, Ann Arbor, Michigan 48109, USA}
\affiliation{International Center for Quantum Materials, School of Physics, Peking University, Beijing 100871, China}
\author{Junbang~Zeng}
\altaffiliation[Present address: ]{Max Planck Institute for Solid State Research, D-70569 Stuttgart, Germany}
\affiliation{International Center for Quantum Materials, School of Physics, Peking University, Beijing 100871, China}
\author{Jiarui~Li}
\altaffiliation[Present address: ]{Department of Physics, Massachusetts Institute of Technology, Cambridge, Massachusetts 02139, USA}
\affiliation{International Center for Quantum Materials, School of Physics, Peking University, Beijing 100871, China}
\author{Izabela~Bialo}
\affiliation{AGH University of Science and Technology, Faculty of Physics and Applied Computer Science, 30-059 Krakow, Poland}
\affiliation{Institute of Solid State Physics, TU Wien, 1040 Vienna, Austria}
\author{Martin~Bluschke}
\affiliation{Max Planck Institute for Solid State Research, D-70569 Stuttgart, Germany}
\affiliation{Helmholtz-Zentrum Berlin f\"{u}r Materialien und Energie, Wilhelm-Conrad-R\"{o}ntgen-Campus BESSY II, 12489 Berlin, Germany}
\author{Yang~Tang}
\affiliation{School of Physics and Astronomy, University of Minnesota, Minneapolis, Minnesota 55455, USA}
\author{Jacob~Freyermuth}
\affiliation{School of Physics and Astronomy, University of Minnesota, Minneapolis, Minnesota 55455, USA}
\author{Guichuan~Yu}
\affiliation{School of Physics and Astronomy, University of Minnesota, Minneapolis, Minnesota 55455, USA}
\author{Ronny~Sutarto}
\affiliation{Canadian Light Source, Saskatoon, Saskatchewan S7N 2V3, Canada}
\author{Feizhou~He}
\affiliation{Canadian Light Source, Saskatoon, Saskatchewan S7N 2V3, Canada}
\author{Eugen~Weschke}
\affiliation{Helmholtz-Zentrum Berlin f\"{u}r Materialien und Energie, Wilhelm-Conrad-R\"{o}ntgen-Campus BESSY II, 12489 Berlin, Germany}
\author{Wojciech~Tabis}
\affiliation{School of Physics and Astronomy, University of Minnesota, Minneapolis, Minnesota 55455, USA}
\affiliation{AGH University of Science and Technology, Faculty of Physics and Applied Computer Science, 30-059 Krakow, Poland}
\affiliation{Institute of Solid State Physics, TU Wien, 1040 Vienna, Austria}
\author{Martin~Greven}
\email[]{greven@umn.edu}
\affiliation{School of Physics and Astronomy, University of Minnesota, Minneapolis, Minnesota 55455, USA}
\author{Yuan~Li}
\email[]{yuan.li@pku.edu.cn}
\affiliation{International Center for Quantum Materials, School of Physics, Peking University, Beijing 100871, China}
\affiliation{Collaborative Innovation Center of Quantum Matter, Beijing 100871, China}

\begin{abstract}
Using resonant X-ray diffraction and Raman spectroscopy, we study charge correlations and lattice dynamics in two model cuprates, HgBa$_{2}$CuO$_{4+\delta}$ and HgBa$_{2}$CaCu$_{2}$O$_{6+\delta}$. We observe a maximum of the characteristic charge order temperature around the same hole concentration ($p \approx 0.09 $) in both compounds, and concomitant pronounced anomalies in the lattice dynamics that involve the motion of atoms in and/or adjacent to the CuO$_2$ layers. These anomalies are already present at room temperature, and therefore precede the formation of the static charge correlations, and we attribute them to an instability of the CuO$_2$ layers. Our finding implies that the charge order in the cuprates is an emergent phenomenon, driven by a fundamental variation in both lattice and electronic properties as a function of doping.

\end{abstract}


\maketitle

The nature of the charge-order phenomena in the high-temperature superconducting cuprates has attracted intense research interest \cite{KeimerNature2015}.
Historically, the tendency toward charge order (CO) in these doped Mott insulators \cite{LeeRMP2006} was first noted theoretically in the context of charge-magnetic stripes \cite{ZaanenPRB1989,KivelsonRMP2003}, which were subsequently observed in the La-214 cuprates near $p=1/8$ doping \cite{TranquadaNature1995,AbbamonteNatPhys2005}.
Related observations made for other cuprate families \cite{WuNature2011,GhiringhelliScience2012,daSilvaNetoScience2014,CominScience2014,TabisNatCom2014,daSilvaNetoScience2015,ChanNatCommun2016,TabisPRB2017}, however, pointed to a connection to Fermi-surface nesting properties of the mobile carriers \cite{daSilvaNetoScience2014,CominScience2014,TabisNatCom2014,daSilvaNetoScience2015}, and indeed the doping dependence of the CO wave vector was found to differ from that of stripes \cite{CominAnnualReview2016}.
In addition to the general observation of two-dimensional short-range charge correlations, recent experiments revealed new phenomena, including re-entrant CO in overdoped Bi$_2$Sr$_2$CuO$_{6+\delta}$ \cite{PengNatMater2018}, three-dimensional CO induced by a variety of external fields in YBa$_2$Cu$_3$O$_{6+\delta}$ \cite{GreberScience2015,ChanNatCommun2016,BluschkeNatcom2018,KimScience2018}, and significant dynamic charge correlations already at high temperatures \cite{Chaix2017,Arpaia2018,BiqiongYu2019}.
These observations indicate an intriguing interplay among electron correlations, band-structure properties and lattice response, and they add to the challenge to understand the phase diagram.

\begin{figure}[hbt!]
\includegraphics[width=3.2in]{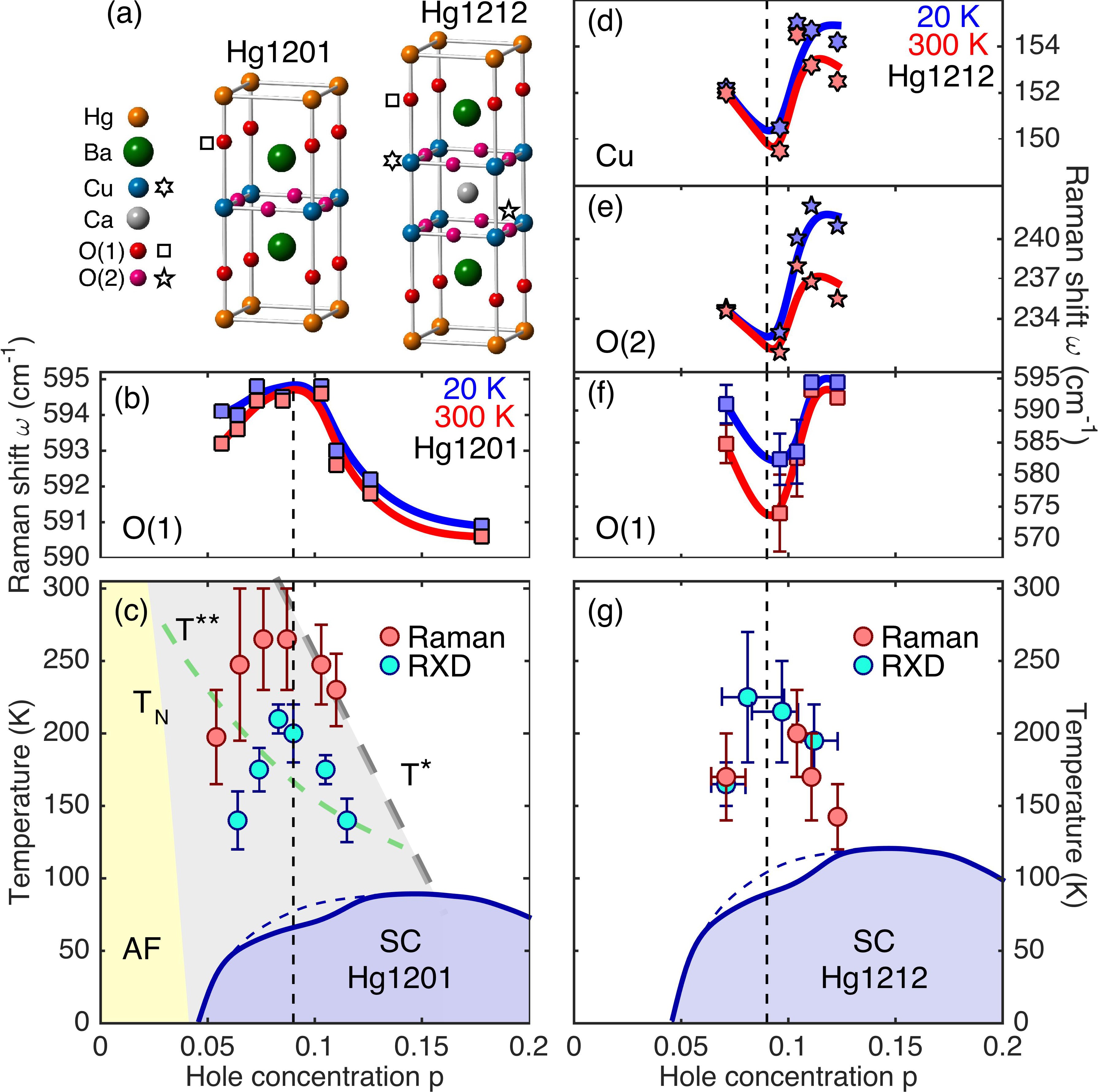}
\caption{\label{Fig1}
(a) Crystal structures of Hg1201 and Hg1212. Empty symbols indicate the primary atoms involved in the $c$-axis vibration of the $A_\mathrm{1g}$ phonons and correspond to (b), (d), (e) and (f). (b) Doping dependence of the apical-oxygen $A_{1g}$ phonon in Hg1201. (c) Phase diagram of Hg1201 \cite{TabisPRB2017}. Symbols indicate characteristic onset temperatures $T_\mathrm{CO}$ of charge correlations determined via Raman spectroscopy and RXD. (d-f) Doping dependence of $A_\mathrm{1g}$ phonons in Hg1212. (g) $T_\mathrm{CO}$ for Hg1212. The doping dependence of $T_\mathrm{c}$ is scaled up from that of Hg1201 \cite{YamamotoPRB2000} to $T_\mathrm{c,max}=127$ K.
}
\end{figure}

The present study aims to investigate the connection between the lattice and incipient CO in HgBa$_{2}$CuO$_{4+\delta}$ (Hg1201) and HgBa$_{2}$CaCu$_{2}$O$_{6+\delta}$ (Hg1212), two cuprates that may be viewed as model systems since they feature simple tetragonal primitive cells (with one and two CuO$_2$ layers, respectively) and the highest optimal critical temperature values ($T_\mathrm{c,max}=$ 98 K and 127 K, respectively) among single- and double-layer cuprates \cite{EisakiPRB2004,WangPRM2018}. Hg1201 has been studied more extensively than Hg1212, and the model nature of the former is also apparent from, \textit{e.g.}, charge transport \cite{BarisicPRB2008,Barisic2013,Barisic2013a,Chan2014,ChanNatCommun2016,Popcevic2018,Pelc2019}, optical spectroscopy \cite{vanHeumen2009,Mirzaei2013,HintonSciRep2016}, and neutron scattering experiments \cite{Li2011,Chan2016a,Chan2016}.

Our key results are summarized in Fig.~\ref{Fig1}. We observe broad maxima of the CO onset temperatures ($T_\mathrm{CO}$) as a function of the hole concentration in the CuO$_2$ layers, using both resonant X-ray diffraction (RXD) and Raman spectroscopy. In both systems, the maximal $T_\mathrm{CO}$ is found in samples with critical temperatures ($T_\mathrm{c}$) about 70\% of the maximal $T_\mathrm{c}$ \cite{SM}, which corresponds to $p \approx 0.09$ \cite{YamamotoPRB2000}. These results confirm previous RXD observations for Hg1201 \cite{TabisNatCom2014,TabisPRB2017} via Raman spectroscopy, extend them to Hg1212 using both techniques, and hence establish the universal existence of CO phenomena in the first two members of the Hg-family of cuprates. Most importantly, our Raman data also reveal a pronounced anomaly in the lattice dynamics of both Hg1201 and Hg1212. This anomaly occurs at about the same doping level as the maximal $T_\mathrm{CO}$, pertains to phonons that primarily involve $c$-axis motion of atoms in and/or adjacent to the CuO$_2$ layers, and is clearly visible already at room temperature, \textit{i.e.}, above $T_\mathrm{CO}$. These results and related earlier Raman scattering work on Hg1201 \cite{LiPRL2013} suggest dual structural and electronic variations at the origin of the $T_\mathrm{CO} (p)$ maximum and the $T_\mathrm{c} (p)$ plateau. Indeed, while there exists evidence for dynamic charge correlations throughout the entire superconducting doping range from neutron scattering measurements of phonon anomalies \cite{ParkPRB2011} and resonant inelastic X-ray scattering (RIXS) measurements \cite{Chaix2017,Arpaia2018,BiqiongYu2019}, the relatively unstable lattice may promote the CO phenomena in this narrower doping range.

\begin{figure}
\includegraphics[width=3.375in]{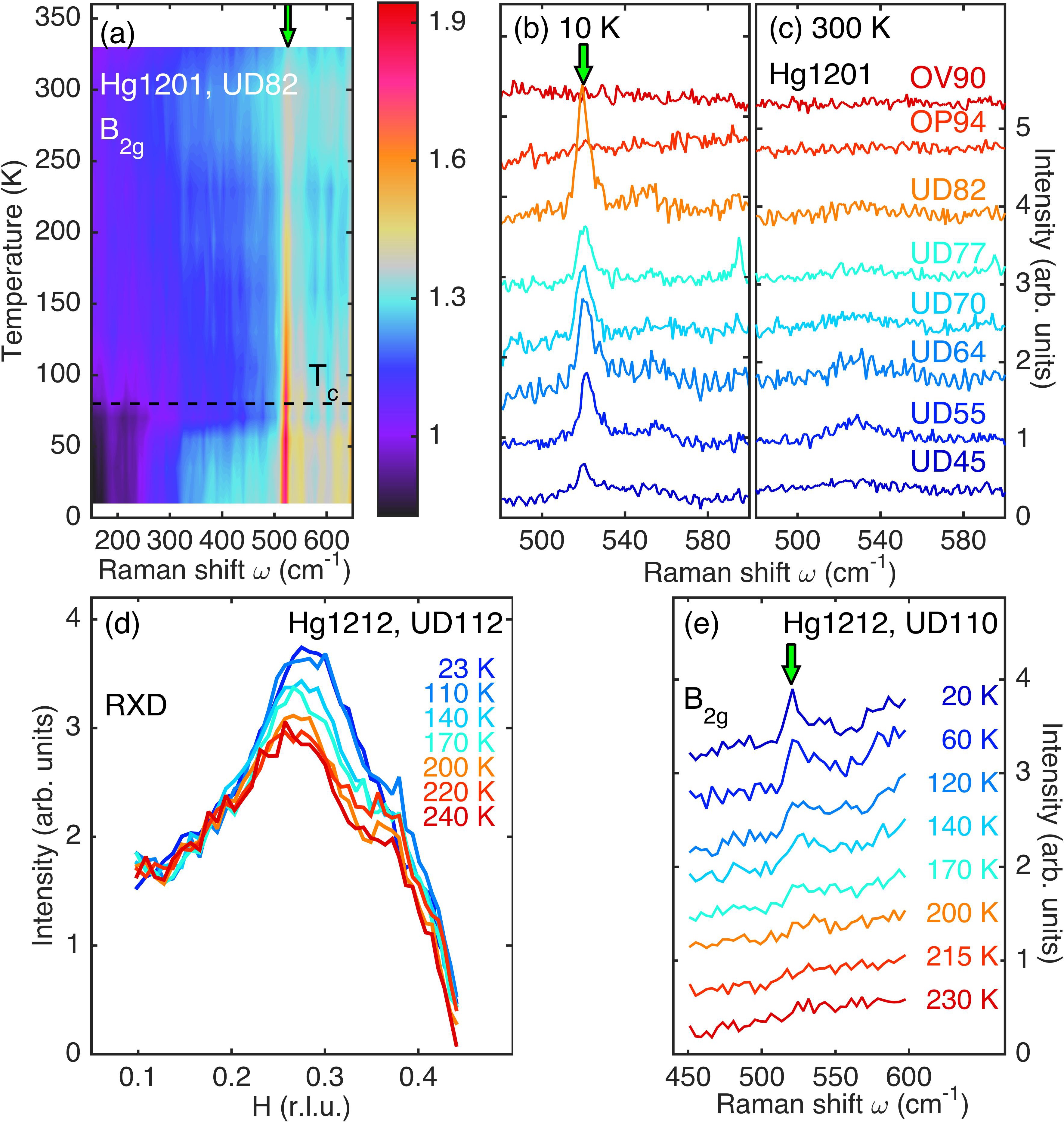}
\caption{\label{Fig2}
(a) Temperature dependence of $B_\mathrm{2g}$ Raman spectra taken on a Hg1201 UD82 sample. The arrow points at 520 cm$^{-1}$, where a sharp phonon peak develops below $T\approx250$ K, whereas a broader peak persists at slightly higher energy above 250 K. (b and c) Doping evolution of the Raman $B_\mathrm{2g}$ phonon peak at 10 and 300 K, respectively, for Hg1201. (d) RXD data taken on a Hg1212 UD112 sample, revealing the development of charge correlation below $T\approx200$ K. (e) Temperature dependence of the Raman $B_{2g}$ phonon peak measured on a Hg1212 UD110 sample. Data in (b), (c), and (e) have been offset for clarity.
}
\end{figure}

The Hg1201 and Hg1212 single crystals were grown with flux methods \cite{ZhaoAdvMater2006,WangPRM2018}. Samples with different oxygen (and hence hole) concentrations were obtained either in the as-grown state or via post-growth annealing \cite{BarisicPRB2008,WangPRM2018}. For Hg1201, we measured underdoped (UD), optimally-doped (OP) and overdoped (OD) samples, and we label samples as, \textit{e.g.}, ``UD82'', where the number indicates $T_\mathrm{c}$ in Kelvin. For Hg1212, all samples were underdoped because annealing crystals to high doping is technically difficult \cite{WangPRM2018}. The RXD experiments were performed at the UE46 PGM-1 beamline of the BESSY-II synchrotron and the REIXS beamline of the Canadian Light Source. The measurements were performed with soft X-ray photons tuned to the energy of the Cu $L_{3}$ resonant absorption edge, on crystals that were freshly polished parallel to the $ab$ plane \cite{SM}. The Raman scattering experiments were performed in a confocal back-scattering geometry with a Horiba Jobin Yvon LabRAM HR Evolution spectrometer equipped with 600 and 1800 gr/mm gratings and a liquid-nitrogen-cooled CCD detector. The $\lambda = 632.8$ nm line from a He-Ne laser was used for excitation. The laser beam had a power of less than 1.0 mW, and was focused down to a $\sim$ 10-$\mu$m-diameter spot onto the freshly-polished samples that were mounted in a liquid-helium flow cryostat and studied under ultrahigh vacuum.

We present two types of Raman spectra, obtained with incident and scattered linear photon polarization (1) parallel to each other and along a diagonal of the CuO$_2$ plaquette, and (2) perpendicular to each other and along the equivalent $a$ and $b$ axes. The two measurement geometries correspond to the $A_\mathrm{1g} + B_\mathrm{2g}$ and the $B_\mathrm{2g}$ representations of the $D_{4h}$ point group \cite{DevereauxRMP2007}, and are denoted as ``$A_\mathrm{1g}$'' and ``$B_\mathrm{2g}$'', respectively.
According to factor-group analysis \cite{RousseauJRamanSpec1981}, neither Hg1201 nor Hg1212 possess Raman-active $B_\mathrm{2g}$ phonons, owing to their high-symmetry structure. Any new phonon peaks that emerge in this geometry therefore indicate a lowering of the symmetry, a natural consequence of CO. The pure $B_\mathrm{2g}$ geometry was used because it features fewer weakly-temperature-dependent defect phonon peaks than the $A_\mathrm{1g}$ and $B_\mathrm{1g}$ counterparts \cite{LiPRL2013,LiPRL2012,GallaisPhysicaC2004}.

Figure~\ref{Fig2}(a-c) displays some of our Raman $B_\mathrm{2g}$ spectra for Hg1201. For the UD82 sample [Fig.~\ref{Fig2}(a)], for example, a sharp and well-defined phonon peak forms below $T\approx250$ K \cite{SM}, somewhat above the CO temperature previously determined \cite{TabisPRB2017} by RXD at a similar doping [Fig.~\ref{Fig1}(c)]. Nevertheless, we regard this peak's temperature dependence as an empirical indication of the formation of charge correlations for several reasons. First, $T_\mathrm{CO}$ is not well-defined for any cuprate, including RXD and Raman scattering results for Hg1201 and Hg1212 \cite{SM,TabisPRB2017}. In particular, the non-monotonic ``background'' in RXD data [Fig.~\ref{Fig2}(d)], which was recently ascribed to dynamic charge correlations in the optical phonon range \cite{BiqiongYu2019}, contributes to the uncertainty in the estimation of $T_\mathrm{CO}$. As an aside, we note here that the characteristic CO temperature obtained for Hg1201 from nonlinear optical measurements is in good agreement with the RXD result \cite{HintonSciRep2016,TabisPRB2017}. Second, despite the systematic offset, $T_\mathrm{CO}$ determined by both Raman \cite{SM} and RXD \cite{TabisPRB2017} exhibits a broad maximum at about $p\approx0.09$ [Fig.~\ref{Fig1}(c)]. Moreover, already at room temperature [Fig.~\ref{Fig2}(c)], the Raman data show a broad ``precursor'' peak at slightly higher energy ($\sim 530$ cm$^{-1}$) for samples with intermediate doping levels, consistent with the doping range over which charge correlations are seen by RXD at lower temperatures. Third, at some doping levels we find a decrease of the Raman peak amplitude below $T_\mathrm{c}$ \cite{SM}, consistent with the notion that CO competes with superconductivity \cite{GhiringhelliScience2012,ChangNatPhys2012}. This is the clearest evidence for such competition in Hg1201, and it renders Raman scattering a sensitive probe of CO in this system, which may in turn explain the higher observed $T_\mathrm{CO}$ compared to the RXD result. Similar competition has been reported for YBa$_2$Cu$_3$O$_{6+\delta}$ \cite{BakrPRB2013}, albeit in a different Raman scattering configuration. Finally, dynamic charge correlations above $T_\mathrm{CO}$ \cite{BiqiongYu2019} may allow disorder to induce static CO-related lattice distortions nearby, which may explain why the Raman effect is observed up to higher $T$ than $T_\mathrm{CO}$ determined by RXD.

Although the Raman $B_\mathrm{2g}$ phonon peak is more difficult to resolve for Hg1212 due to its weak intensity, an even better correspondence between $T_\mathrm{CO}$ values determined by RXD and Raman is observed [Figs.~\ref{Fig1}(g) and \ref{Fig2}(d and e)]. A possible physical origin of this difference might be that the two probes have different degrees of sensitivity to static versus dynamic correlations, and that dynamic correlations above $T_\mathrm{CO}$ are weaker and/or static correlations below $T_\mathrm{CO}$ stronger in Hg1212.
Similar to Hg1201, the characteristic wave vector of charge correlations in Hg1212 is determined by RXD to be about 0.3 reciprocal lattice units (r.l.u.) [Fig.~\ref{Fig2}(d)], and the correlation length is merely a few lattice constants \cite{SM}. By assuming a similar $T_\mathrm{c}$ versus $p$ relationship as for Hg1201 \cite{YamamotoPRB2000}, but with the maximal $T_\mathrm{c}$ scaled to 127 K \cite{SchillingNature1993,WangPRM2018}, we arrive at a $T_\mathrm{CO}$ dome for Hg1212 which is also centered at $p\approx0.09$ [Fig.~\ref{Fig1}(g)]. Altogether, these results indicate the universal presence of charge correlations in the Hg-family of cuprates. The short-range nature of these correlations, as well as the somewhat gradual increase of the intensity on cooling (rather than the development of a true order parameter), imply considerable nanoscale inhomogeneity \cite{TorreNatPhys2016,TorrePRB2016,TorreNewJPhys2015,YuePreprint,TabisPRB2017,WuNatCommun2015}, which is the result of dopant-related point disorder \cite{EisakiPRB2004} and/or inherent structural and electronic inhomogeneity \cite{Pelc2018,Pelc2019a,Pelc2019c}.

\begin{figure}
\includegraphics[width=3.375in]{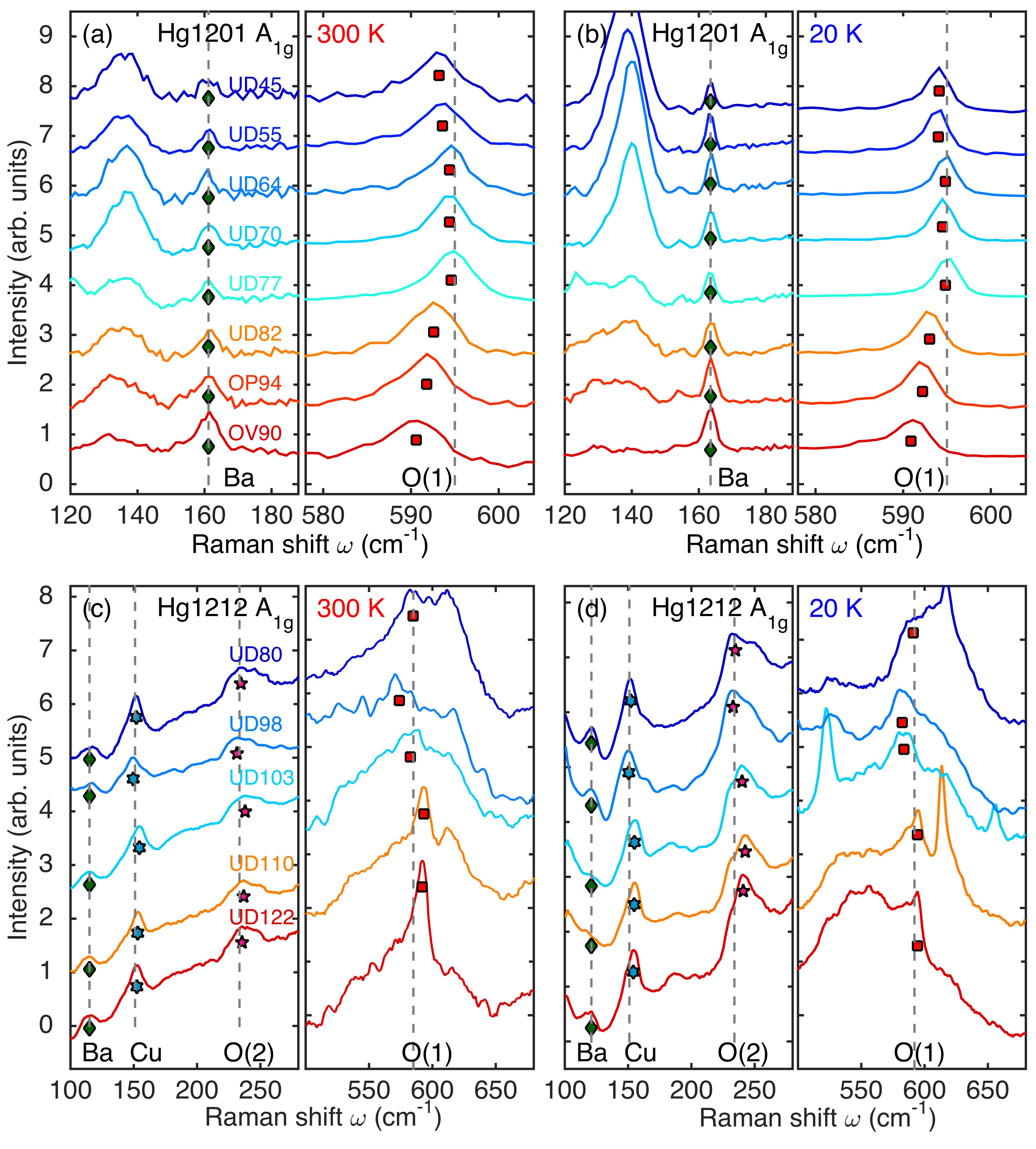}
\caption{\label{Fig3}
(a, b) Doping dependence of $A_{1g}$ Raman peaks in Hg1201 at 300 and 20 K, respectively. (c, d) Doping dependence of $A_{1g}$ Raman peaks in Hg1212 at 300 and 20 K, respectively. Data are offset for clarity. The atoms primarily involved in the vibrations are indicated at the bottom. Peak positions are indicated by symbols consistent with those in Fig.~\ref{Fig1}. Vertical dashed lines are guides to the eye.
}
\end{figure}

We now turn to another major aspect of our findings, namely the structural instability near $p\approx0.09$. This is best seen from Raman measurements in $A_\mathrm{1g}$ geometry, which detect two and four Brillouin zone-center vibrational modes in Hg1201 and Hg1212 \cite{WangPRM2018}, respectively, in addition to defect-related and multi-phonon peaks. Here we focus on the relatively sharp and $A_\mathrm{1g}$-symmetry-allowed phonon modes, which involve $c$-axis motion of atoms [Fig.~\ref{Fig1}(a)]. We first notice that the lowest-energy mode, marked by diamond symbols in Fig.~\ref{Fig3}, has a doping-independent frequency in both Hg1201 and Hg1212. Following previous understanding of the lattice dynamics \cite{WangPRM2018,ZhouPhysicaC1996,BauerJPhysCondensMatter2010}, we attribute this mode primarily to the vibration of the heavy barium atoms. This assignment is very reliable for Hg1201, because only the apical oxygen and the barium atoms can participate in $A_\mathrm{1g}$ modes, and their masses are very different. The corresponding mode in Hg1212 may also involve some $c$-axis motion of the CuO$_5$ pyramids as rigid units, which could explain the mode's lower frequency compared to Hg1201. The doping independence of these modes is not surprising, since the change in the oxygen content is minute and confined to the Hg layer \cite{EisakiPRB2004}, and because the modes do not deform the CuO$_2$ layers.

In contrast, we find that \textit{all} of the remaining $A_\mathrm{1g}$ phonons exhibit a pronounced frequency anomaly with doping, around UD77 in Hg1201 and UD98 in Hg1212. This is summarized in Fig.~\ref{Fig1}(b) and Fig.~\ref{Fig1}(d-f), respectively. The O(1) phonon in Hg1201 exhibits a frequency maximum, whereas the Cu, O(1) and O(2) phonons in Hg1212 exhibit frequency minima. While the different manifestation of the anomaly in the two systems warrants further study, all these effects are observed at $p \approx 0.09$, \textit{i.e.}, where $T_\mathrm{CO}$ exhibits a maximum in both systems. On the technical side, we remark that the effect is very prominent in Hg1201, where we expect phonons that involve the in-plane Cu and O(2) atoms to exhibit related anomalous doping dependence as well, and it will be interesting to study those modes by, \textit{e.g.}, optical spectroscopy (those modes are not Raman-active in Hg1201), and to search for similar effects in other cuprates. Although the anomalous phonon peaks in Hg1212 are less well separated from neighboring defected-related peaks, the three peaks exhibit the same anomalous trend with doping [albeit with different frequency variations - note the different vertical scales in Fig.~\ref{Fig1}(d-f)]; we made sure that the absolute frequency reading from the spectrometer is reliable by performing frequent spectral calibration with a neon lamp.

As seen from Fig. 1, the observed anomalies are similarly pronounced at 20 K and 300 K, which indicates that they are not a consequence of the CO, but rather a precursor phenomenon.  It is revealing that all of the anomalous phonons bring about lattice deformations that could change the electronic properties in the quintessential CuO$_2$ layers: the apical-oxygen height is known to affect $T_\mathrm{c}$ \cite{PavariniPRL2001} and magnetic exchange interactions \cite{PengNatPhys2017}, because it affects the electronic in-plane hopping range, whereas the relative $c$-axis displacements of in-plane Cu and O(2) atoms are expected to influence the band structure and the nearest-neighbor exchange interactions even more strongly, since they modify the Cu-O-Cu bond angle. Therefore, we believe that the phonon anomalies actually indicate a fundamental variation in the electronic properties of the CuO$_2$ layers near $p\approx0.09$. In a previous electronic Raman scattering study of Hg1201 \cite{LiPRL2013}, it was found that the size of the antinodal superconducting gap exhibit pronounced variations around the same doping level, in support of a correlation between the electronic and lattice degrees of freedom.

We believe that it is important to understand the microscopic origin of the electronic variations near $p\approx0.09$, since the doping level coincides with the center of the dome-shaped CO phase and $T_\mathrm{c} (p)$ plateau, and because the charge-density correlations are related to other important aspects of high-$T_\mathrm{c}$ physics including the pseudogap and superconductivity itself \cite{KeimerNature2015,BiqiongYu2019,LoretNatPhys2019}. Since the CO and $T_\mathrm{c} (p)$ plateau are universally observed in the cuprates, it is likely that the phenomena observed here for Hg1201 and Hg1212 reflect a universal underlying electron-lattice instability of the doped CuO$_2$ layers, with somewhat different manifestations for different compounds. For example, the La-214 cuprates exhibit a structural instability near $p=1/8$ where charge-spin stripe correlations are most pronounced, along with a suppression of $T_\mathrm{c}$ \cite{Fujita2012}. The stripe formation can be understood as a tendency of the doped CuO$_2$ layers to maintain local electronic environments that resemble the undoped Mott-insulating state. Another example is YBa$_2$Cu$_3$O$_{6+\delta}$ (YBCO), for which the $T_\mathrm{c} (p)$ plateau is centered at $p=1/8$ \cite{Liang2006}, where CO is most pronounced \cite{WuNature2011,GhiringhelliScience2012}. The different apparent doping levels at which this instability occurs ($p\approx0.09$ for Hg1201 and Hg1212 versus $p\approx 1/8$ for La-214 and YBCO) might be associated with material-specific properties and the indirect manner in which the hole concentration is estimated \cite{YamamotoPRB2000,TabisPRB2017}.

Our finding points to the importance to understand the role of disorder \cite{TorreNatPhys2016,TorrePRB2016,TorreNewJPhys2015,YuePreprint,TabisPRB2017,WuNatCommun2015} in the formation of short-range CO: if there exists a fundamental combined electronic-lattice instability, it is likely amplified by disorder effects. We note that recent work on oxide superconductors, including the cuprates, points to universal lattice (and hence electronic) inhomogeneity in perovskite-based compounds \cite{Pelc2019c}, i.e., there exist substantial deviations from the average lattice structure that may be enhanced at certain (effective) doping levels. In the case of the cuprates, the primary CO phenomenon is the localization of one hole per CuO$_2$ with decreasing doping and temperature \cite{Pelc2019a}, and it will be important to understand the anomalies observed in the present work in this larger context.

In summary, we report a comprehensive spectroscopic study of the first two members of the Hg-family of cuprates. We observe common behavior in both the dome-like shape of the CO region in the phase diagram and the presence of a phonon-frequency anomaly with doping, which is already observable for temperatures above $T_\mathrm{CO}$. The occurrence of these behaviors around the same doping level, along with the fact that they both involve the CuO$_2$ layers, suggests that they are fundamentally related. Our finding calls for closer scrutiny of combined lattice and electronic anomalies in other high-$T_\mathrm{c}$ cuprates in order to further clarify how the CO phenomena and superconductivity are influenced by doping-dependent variations that occur at high energy and temperature scales.

\begin{acknowledgments}
We wish to thank M. Le Tacon, B. Keimer, D. Pelc, and Y. Y. Peng for discussions, and Enrico Schierle for assistance during the experiments at HZB. The work at Peking University was supported by the National Natural Science Foundation of China (grant nos. 11874069 and 11888101) and Ministry of Science and Technology of China (grant Nos. 2018YFA0305602 and 2015CB921302). The work at the University of Minnesota was funded by the Department of Energy through the University of Minnesota Center for Quantum Materials under DE-SC-0016371. The work at TU Wien was funded by the European Research Council (ERC Consolidator Grant No 725521). Part of research described in this paper was performed at the Canadian Light Source, which is supported by the Canada Foundation for Innovation, Natural Sciences and Engineering Research Council of Canada, the University of Saskatchewan, the Government of Saskatchewan, Western Economic Diversification Canada, the National Research Council Canada, and the Canadian Institutes of Health Research.
\end{acknowledgments}

\preprint{Preprint}
\bibliography{reference}
\end{document}